\documentstyle[aps,multicol,tighten,epsfig]{revtex}

\begin{document} 
\draft

\title{Chaos synchronization in long-range coupled map lattices}                
 
\author{C. Anteneodo$^{1}$,  A.M. Batista$^{2}$ and R.L. Viana$^{3}$} 
   
\address{1. Centro Brasileiro de Pesquisas F\'{\i}sicas,\\ 
            Rua Dr. Xavier Sigaud 150, 22290-180, 
            Rio de Janeiro, RJ, Brazil; \\ 
         2. Departamento de Matem\'atica e Estat\'{\i}stica,            
            Universidade Estadual de Ponta Grossa,\\
            84030-900, Ponta Grossa, PR, Brazil;          \\
         3. Departamento de F\'{\i}sica, 
            Universidade Federal do Paran\'a,  \\
            81531-990, Curitiba, PR, Brazil.} 
 
%\date{\today} 
\maketitle 
 
\begin{abstract} 
We investigate the synchronization phenomenon in coupled chaotic map lattices 
where the couplings decay with distance following a power-law.
Depending on the lattice size, the coupling strength and the 
range of the interactions, complete chaos synchronization may be attained.
The synchronization domain in parameter space can be analytically delimited 
by means of the condition of negativity of the largest transversal Lyapunov 
exponent. 
Here we analyze in detail the role of all the system parameters in 
the ability of the lattice to achieve complete synchronization, 
testing analytical results with the outcomes of numerical experiments.
\end{abstract} 
 
 
% 05.45.Ra :    Coupled map lattices                         
% 05.45.-a :    Nonlinear dynamics 
% 05.45.Xt :    Synchronization; coupled oscillators                    
 
\begin{multicols}{2} 

\narrowtext 

%\section*{Introduction}

Coupled map lattices (CMLs), dynamical systems with discrete 
space and time, are being intensively investigated nowadays as models 
of spatiotemporal phenomena occurring in a wide variety 
of systems of physical, biological and technical interest \cite{kanekobook}. 
In this letter we will deal with the phenomenon of synchronization and, in 
particular, amongst the various kinds of synchronized behavior, 
with  the {\em complete synchronization} (CS) \cite{boccaletti} occurring 
in CMLs with regular long-range interactions. 
Most work done so far on synchronization in CMLs has focused on 
two extreme coupling types: local (nearest-neighbor)\cite{nneighbor} 
and global ("mean field") ones \cite{meanfield}.  
However, non-local couplings are relevant to a variety of situations 
ranging from neural networks \cite{neural} to physico-chemical reaction 
systems \cite{chemreac}.

We consider chains of $N$ coupled one-dimensional (1D) chaotic 
maps $x \mapsto f(x)$ whose evolution is given by \cite{pinto00}    
 
\begin{equation} 
x^{(i)}_{n+1}=(1-\varepsilon)f(x^{(i)}_{n})+\frac{\varepsilon} 
{\eta}\sum^{N'}_{r=1} 
\frac{f(x^{(i-r)}_{n})+f(x^{(i+r)}_{n})}{r^{\alpha}}, 
\label{CML} 
\end{equation} 
where $x^{(i)}_n$ represents the state variable for the site 
$i$ $(i=1,2,...,N)$ at time $n$, $\varepsilon\geq 0$ is the 
coupling strength, $\alpha \ge 0$ controls the effective range of 
the interactions and $\eta=2 \sum^{N'}_{r=1}r^{-\alpha}$ is a normalization  
factor, with $N'=(N-1)/2$ for odd $N$.  
The main interest in this coupling scheme resides in the fact that it allows to 
investigate the role of the range of the interactions, scanning from the local 
($\alpha \to\infty$) to the global ($\alpha=0$) cases \cite{viana98}. 

CS takes place when the dynamical variables that define the state of 
each map adopt the same value for all the coupled maps at all times $n$, i.e, 
$x^{(1)}_n=x^{(2)}_n=\ldots=x^{(N)}_n\equiv x^{(*)}_n$.   
Depending on the lattice size and on the range of the interactions, 
there may exist an interval of values of the coupling strength $\varepsilon$, 
for which such state is spontaneously attained, as we have analytically shown 
in previous work \cite{apbv03}. 
It is our purpose here to scrutinize the role of all the 
system parameters in the ability of the lattice to synchronize. 
Analytical results will be compared with the outputs of numerical simulations 
performed for diverse chaotic maps.

%%\section*{Complete synchronization domains}

Complete synchronization can be characterized by a complex 
order parameter defined, for time $n$, as 
$R_n=| \frac{1}{N} \sum_{j=1}^{N} e^{2\pi i x_n^{(j)}}|$ \cite{kuramotobook}. 
Typically a time-averaged amplitude $\bar R$ is computed over a time interval 
long enough to allow the lattice attain the asymptotic state.  
In the CS state, one has $\bar R =1$ within a small allowed deviation. 

Another  diagnostic of complete synchronization can be obtained from 
the Lyapunov spectrum (LS) of the CS states. 
If the chaotic maps are completely synchronized, 
the maximal Lyapunov exponent, in the direction parallel to the 
synchronization manifold (SM), is strictly positive.  
The negativity of the second largest Lyapunov exponent,  
which belongs to the direction transversal to the SM, 
indicates the stability of the synchronized state 
under small transversal displacements \cite{gade}. 

In our case the lattice dynamics given by Eq. (\ref{CML}) can be written  as
$x^{(i)}_{n+1} = \sum_j F_{ij} f(x^{(j)}_{n})$, 
where ${\bf F}$ is a matrix of the form 

\begin{equation} 
{\bf F}=\biggl[(1-\varepsilon)\openone+\frac{\varepsilon} 
{\eta}{\bf B} \biggr], 
\label{CMLmatrix} 
\end{equation} 
with $\openone$ the $N\times N$ identity  matrix and ${\bf B}$ defined by 
$B_{jk}=  (1-\delta_{jk})/r^{\alpha}_{jk}$,  
being $r_{jk}=\mbox{min}_{l\in \cal{Z}}|j-k+lN|$. 
 
The Lyapunov spectrum is obtained from the dynamics of tangent vectors 
$\xi$,  which in turn is obtained by differentiation 
of the original evolution equations.
In matrix form the tangent dynamics reads 
$\xi_n = {\bf \cal T}_n\xi_0$, 
where ${\bf \cal T}_n $ is product of $n$ Jacobian matrices 
calculated at successive points of a given trajectory.
If $\Lambda^{(1)},\ldots ,\Lambda^{(N)}$ are the eigenvalues of 
$\hat{\Lambda}= {\displaystyle \lim_{n\rightarrow \infty}} 
({\cal T}^{T}_n {\cal T}_n)^{\frac{1}{2n}}$, the Lyapunov exponents  
are obtained as  $\lambda^{(k)}=\ln\Lambda^{(k)}$, 
for $k=1,\ldots,N$ \cite{ruelle}. 
For the CS states, 
one arrives at the following 
expression for the Lyapunov spectrum \cite{apbv03}
 
\begin{equation} \label{lambdak}
\bar\lambda^{(k)}\,=\, \lambda_{U} \,+\, \ln \left| 1-\varepsilon 
+\varepsilon \frac{b^{(k)}}{\eta} \right|, 
\end{equation}
where $\lambda_{U}>0$ is the Lyapunov exponent of the uncoupled chaotic map, and 
$b^{(k)}$ are the eigenvalues of ${\bf B}$ that can be obtained by Fourier 
diagonalization and,  for odd $N$, read 

\begin{equation} 
b^{(k)}=2\sum^{N^{\prime}}_{m=1}\frac{\cos(2\pi k m/N)}{m^{\alpha}}, 
\;\;\; 1\le k\le N. 
\label{beigenvalues} 
\end{equation} 
For even $N$, summations run up to $N'=N/2$ and half of the $N'$th term has to 
be substracted. 
The maximal eigenvalue is $b^{(N)}=\eta$ and the minimal one is $b^{(N')}$. 
Except for the cases $k=N'$, with even $N$, and $k=N$, the remaining 
eigenvalues are two-fold degenerate, being $b^{(k)}= b^{(N-k)}$.

In the calculation of Lyapunov exponents $\lambda^{(k)}$,  
notice that the parameters that define the particular  
uncoupled map affect only $\lambda_U$, while the second term  in 
Eq.~(\ref{lambdak}) is determined by the particular cyclic dependence on 
distance in the regular coupling scheme (a power law in our case).
It can be easily verified that, for arbitrary $\alpha$,  
the CS state lies along  
the direction of the eigenvector associated to the largest exponent 
$\bar\lambda^{(N)}$. 
Therefore, the CS state will be transversally 
stable if the $(N-1)$ remaining exponents are negative,  
that is $|1-\varepsilon+\varepsilon b^{(k)}/\eta|< {\rm e}^{-\lambda_U}$, 
$\forall k\neq N$. This is equivalent to requiring that 
the second largest (or largest transversal) 
asymptotic exponent, denoted by  $\bar\lambda^\perp$, be negative.  
This exponent is obtained from Eq.~(\ref{lambdak}) with either $k=1$ 
(hence also $k=N-1$ due to degeneracy) or with $k=N'$ 
(hence also $k=N'+1$ if $N$ is odd), depending on whether 
$|1-\varepsilon +\varepsilon b^{(1)}/\eta|$ is, respectively, 
greater or smaller than $|1-\varepsilon +\varepsilon b^{(N')}/\eta|$. 
The condition  $\bar\lambda^\perp< 0$ leads to 
$\varepsilon_c<\varepsilon <\varepsilon^\prime_c$ 
\cite{apbv03} (see also \cite{rd02}), where 

\begin{eqnarray} 
\label{critical1} 
\varepsilon_c(\alpha,N,\lambda_U) &=&  ( 1-{\rm e}^{-\lambda_U} ) 
 \biggl(1- \frac{b^{(1)}}{\eta} \biggr)^{-1}\;\;\;\mbox{and} \\ 
 \label{critical2}
\varepsilon^\prime_c(\alpha,N,\lambda_U)  &=& 
 ( 1+{\rm e}^{-\lambda_U} )
 \biggl(1- \frac{b^{(N')}}{\eta}\biggr)^{-1} .
\end{eqnarray} 

\begin{figure}[htb]  
\begin{center}
\includegraphics*[bb=42 139 552 659, width=0.4\textwidth]{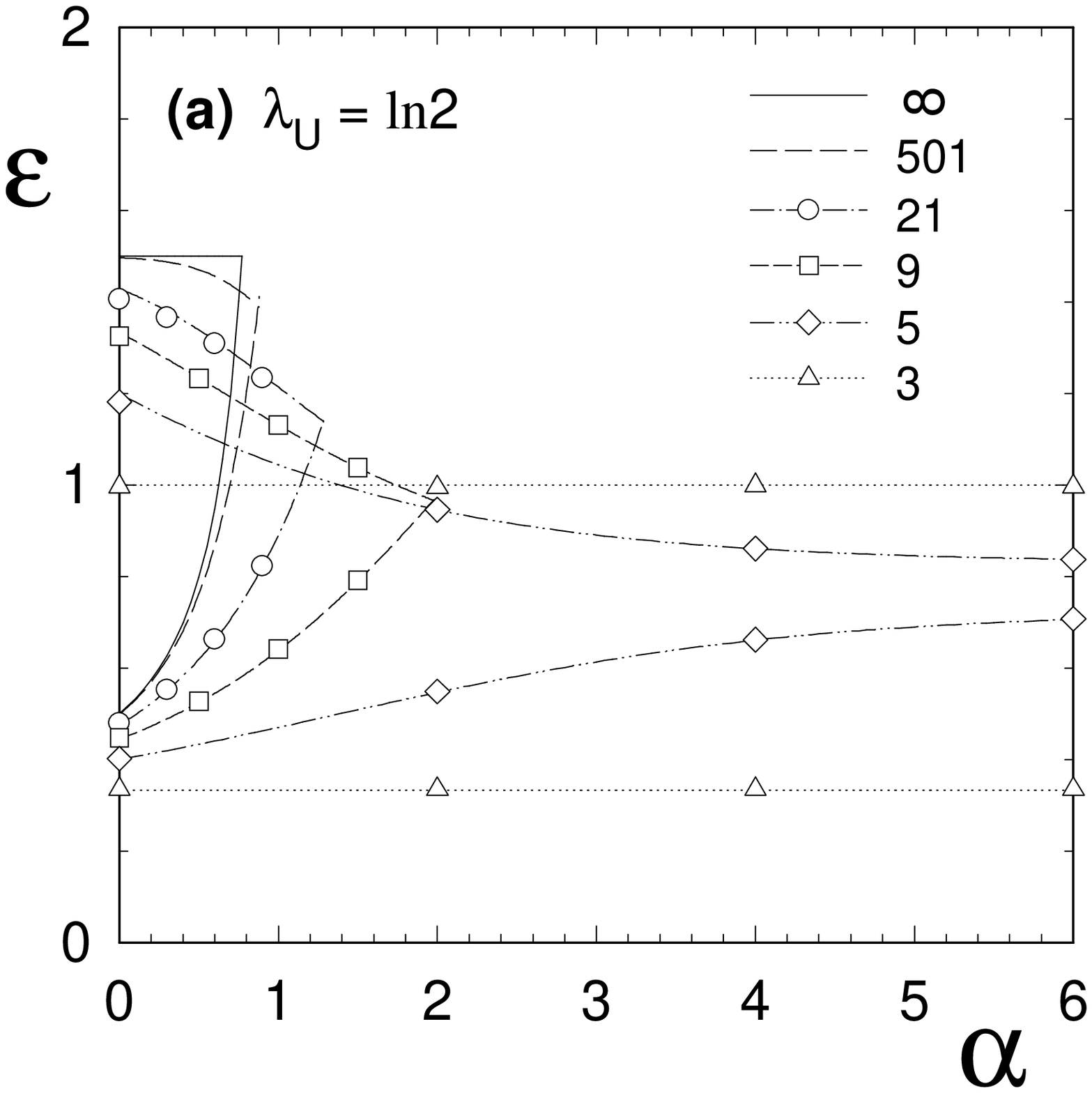}
\includegraphics*[bb=42 139 552 659, width=0.4\textwidth]{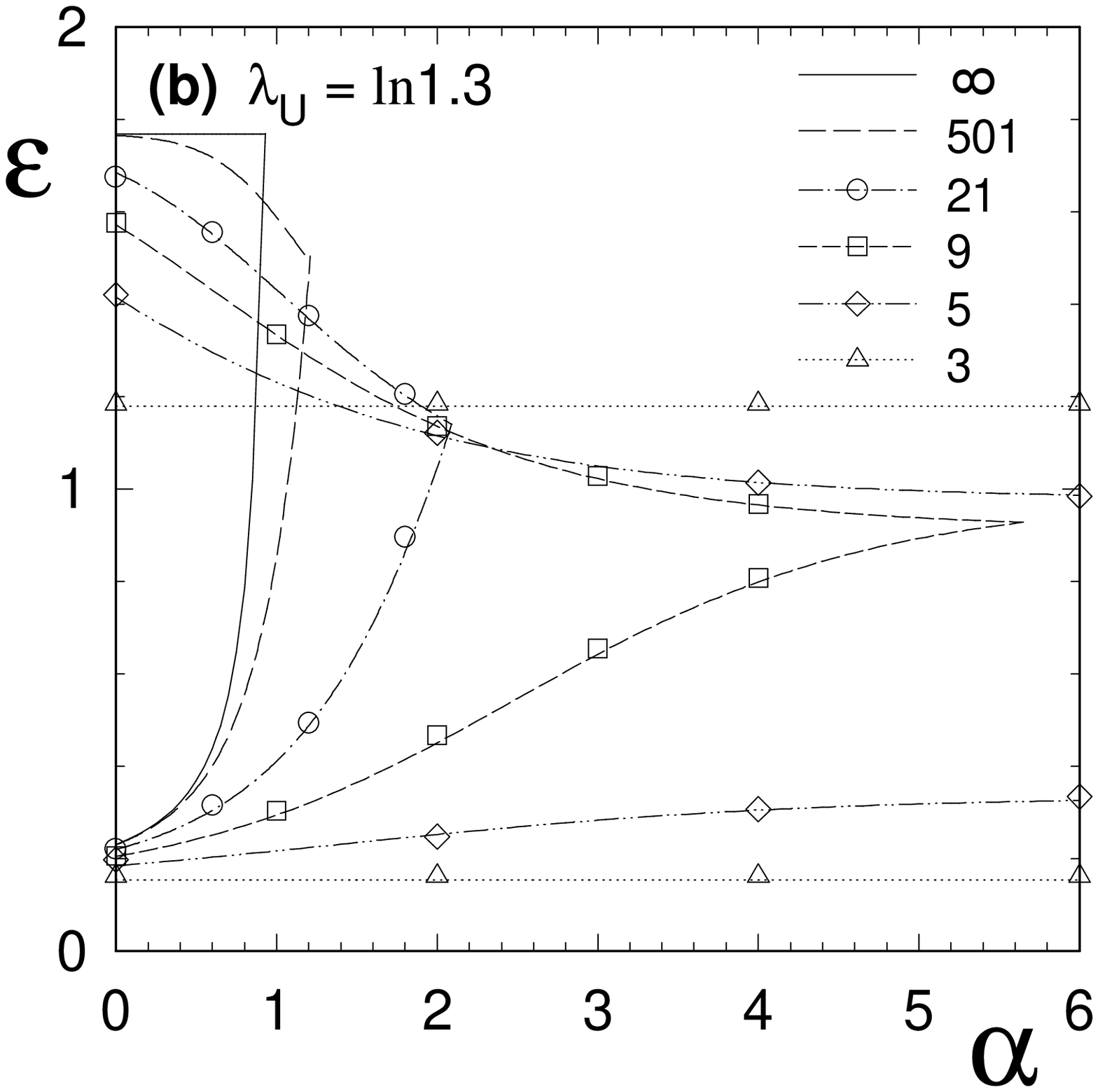}
\end{center} 
\vspace*{-5mm} 
\caption{Synchronization diagram in parameter space ($\alpha,\varepsilon$), for 
different values of $N$ and $\lambda_U= \ln 2$ (a), $\ln 1.3$ (b). 
Lines correspond to analytical predictions;    
symbols to numerical simulations using the Bernoulli (a) or   
triangular (b) maps. 
Synchronization is transversally stable in the  region between the couple of curves 
for each set of values of the parameters.   
} 
\label{fig1} 
\end{figure}

In Fig.~\ref{fig1} we show a variety of critical curves  
in parameter space ($\alpha,\varepsilon$) obtained for different lattice sizes.  
Stable CS states dwell in the region bounded by the $\alpha = 0$ axis, 
and two curve segments. 
The critical curves were obtained analytically from 
Eqs.~(\ref{critical1}) (lower curve) and (\ref{critical2}) (upper curve). 
The symbols shown stand for the numerical results determined from the 
condition $\bar R=1$ with a tolerance of $10^{-6}$.  
Two different values of $\lambda_U$ were considered. 
Numerical results shown in Fig.~\ref{fig1} were computed for the piecewise linear 
(a) Bernoulli shift $f(x)=2x$ (mod 1) (therefore $\lambda_U=\ln 2 $) 
and (b) triangular map \cite{beck} 

\begin{equation}
f_w(x)= 
\left\{ \matrix{    x/w          &  \mbox{for $0\le x\le w$} \cr
                  (1-x)/(1-w)     &  \mbox{for $w< x\le 1$}       } \right.
\end{equation}

\vspace*{5mm}

\noindent
that for $w=0.074$ yields $\lambda_U\simeq \ln 1.3$ 
(notice that $\lambda_U=-w\ln w -(1-w)\ln[1-w]$). 
Additional tests (results not shown here) were performed for other maps such 
as the logistic map $f(x)=\mu x(1-x)$ with $\mu=4$
(hence $\lambda_U=\ln 2$) and $\mu=3.6533$ (hence $\lambda_U\simeq\ln 1.30$) 
yielding the same degree of agreement. 
For these interval maps, in principle, one must have $0\le\varepsilon\le 1$ 
in order to guarantee that the 
state variables $x_n^{(j)}$ will remain inside the interval $[0,1]$. But,   
reinjection into the interval can be performed through an operation, 
for instance (mod 1), such that it does not spoil the Lyapunov exponent 
of the chaotic uncoupled map. 
If trajectory points were not reinjected, one can still look at our results 
as valid for trajectories or trajectory segments as long as the state variables 
remain confined within the given interval. 
Anyway, for other maps such as $f(x)=\exp\{-[(x-0.5)/\sigma]^2\}$\cite{piro}, 
one may  have any coupling $\varepsilon\ge 0$ since the map is naturally 
defined in the full real axis. 
Tests performed with this Gaussian map (results not shown in this letter) 
are also in good accord with theoretical predictions.

In general terms, we observe that for weak coupling, the maps do not synchronize. 
As the coupling strength increases, 
synchronization can occur depending on the lattice parameters $(\alpha,N,\lambda_U)$. 
However, a too high coupling intensity $\varepsilon>\varepsilon^\prime_c$ 
has a destabilizing influence on the CS state and the lattice no longer synchronizes.

Concerning lattice size, already Fig.~\ref{fig1} exhibits the intuitive  
fact that it is more difficult to synchronize 
a larger lattice than a shorter one, all other parameters being kept fixed. 
In the limit $N\to\infty$ we obtain 
 
\begin{equation} 
\varepsilon_c(\alpha,\infty,\lambda_U) \;=\;  
\frac{1- {\rm e}^{-\lambda_U} }{1-C(\alpha)}, 
\label{e1inf} 
\end{equation} 
where $C(\alpha)=\lim_{N\to\infty} b^{(1)}/\eta$.
This limit is equal to one for $\alpha >1$,   
so that Eq.~(\ref{e1inf}) yields a divergent result. For  
$\alpha$ outside the domain of convergence of the series, i.e. $\alpha<1$, one gets 
 
\begin{equation} \label{c1inf}
C(\alpha)\;=\;   
\frac{1-\alpha}{\pi^{1-\alpha}} \int_0^\pi \frac{\cos(x)}{x^\alpha}\,dx . 
\end{equation} 
In that same range of $\alpha$ one has 

\begin{equation} \label{e2inf}
\varepsilon^\prime_c(\alpha,\infty,\lambda_U) \;=\;  1+ {\rm e}^{-\lambda_U},   
\end{equation} 
which is independent on $\alpha$, thus it yields 
a straight line in the plots of Fig.~\ref{fig1}. 
From the intersection of $\varepsilon_c(\alpha,\infty,\lambda_U)$ 
with $\varepsilon^\prime_c(\alpha,\infty,\lambda_U)$ 
it results a critical value of the interaction range $\alpha_c$, such that,   
for $\alpha \le \alpha_c<1$ 
($\alpha_c<d$ in the $d$-dimensional case \cite{apbv03}), synchronization 
is possible even in the thermodynamic limit 
$N\to\infty$ for an appropriate window of $\varepsilon$. 
Observe the corresponding domains in Fig.~\ref{fig1}.

As $N$ diminishes, the upper curve in Fig.~\ref{fig1}, 
which is a straight line for infinite $N$, gains a negative inclination and 
extends for large $\alpha$ to values of $\varepsilon$ less than unity, indicating 
that the destabilizing effect of very strong coupling is more easily attained. 
The lower curve segment has a positive inclination, connecting to the upper 
curve at a point that forms a cusp  for small $N$. For still smaller $N$ 
(e.g., $N\leq 5$ for $\lambda_U=\ln 2$ and $N\leq 8$ for $\lambda_U=\ln 1.3$)
the two critical curves do not intersect each other 
even in the limiting case of first neighbors ($\alpha\to\infty$).

\begin{figure}[htb]  
\begin{center}
\includegraphics*[bb=40 150 550 650, width=0.4\textwidth]{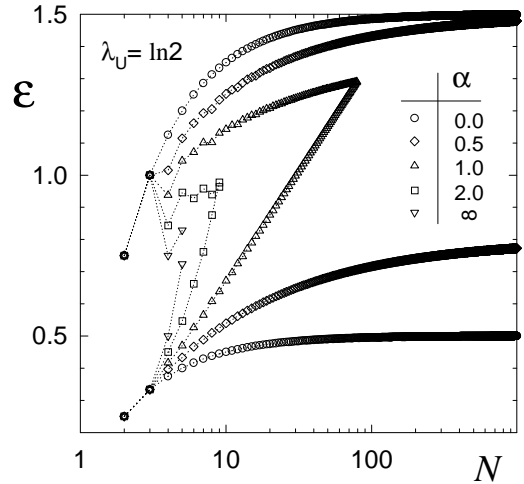}
\end{center}
\vspace*{-2mm}
\caption{\protect Synchronization critical lines in the plane  
$(N,\varepsilon)$ for different values of $\alpha$ and $\lambda_U=\ln 2$.  
In this case $\alpha_c\approx 0.77$. 
Symbols correspond to theoretical calculations. 
Dotted lines are guides to the eyes.  
Synchronization is transversally stable in the  region between the two curves 
for each set of values of the parameters.
} 
\label{fig:Ne} 
\end{figure}

The effect of lattice size can also be observed in 
 Fig.~\ref{fig:Ne} that exhibits the synchronization domains in the plane  
($N,\varepsilon$) for different values of $\alpha$. 
Similar plots have been observed in scale-free networks \cite{ieee03}, as 
if there were an average or effective $\alpha$ in such cases.
While for $\alpha>\alpha_c$ there is un upper bound $N_b(\alpha,\lambda_U)$ of the number of maps 
for which synchronization occurs; for $\alpha<\alpha_c$ any number of maps synchronize 
(because the critical curves in  Fig.~\ref{fig:Ne} do  not intersect). 
Generically it is easier to synchronize a small number of maps. 
Consistently with this observation, lattices of small size 
(e.g., $N\le 5$ for $\lambda_U=\ln 2$ ) can synchronize for 
any $\alpha$, for a certain window of $\varepsilon$ that narrows with 
increasing $\alpha$. For $N\leq3$ there is, naturally, no dependence on $\alpha$ 
and the lattice syncronizes for any $\lambda_U>0$.

\vspace*{5mm}

\begin{figure}[htb] 
\begin{center} 
\includegraphics*[bb=60 120 570 650, width=0.4\textwidth]{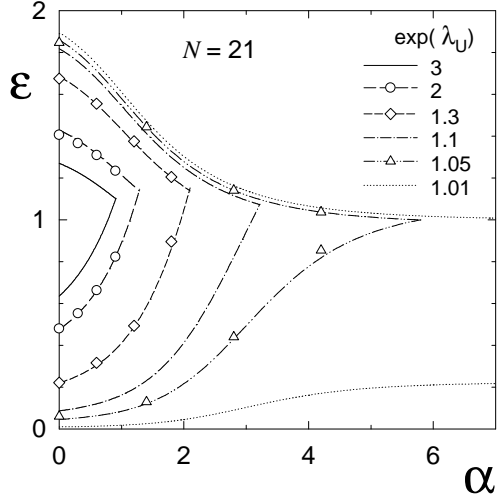}
\end{center} 
\vspace{-5mm} 
\caption{\protect Synchronization domains in parameter plane  
($\alpha,\varepsilon$) for $21$ coupled chaotic maps and various 
values of $\lambda_U$.  
Numerical simulations were performed  for Bernoulli or triangular maps.
Critical curves were obtained analytically from 
Eqs.~(\ref{critical1}) and (\ref{critical2}). 
} 
\label{fig:chaoticity} 
\end{figure}

Although Figs.~\ref{fig1}(a) and \ref{fig1}(b) yield qualitative similar results, 
their comparison 
makes clear that,  as expected, the more chaotic the uncoupled maps are, 
the more difficult becomes to obtain their synchronization. 
The influence of the Lyapunov exponent $\lambda_U$ on the synchronization domains 
in the parameter space ($\alpha,\varepsilon$)  is displayed in Fig.~\ref{fig:chaoticity}.
The exhibited numerical results were acquired for $N=21$ 
either Bernoulli or triangular maps.

\begin{figure}[htb]  
\begin{center}
\includegraphics*[bb=40 210 550 620, width=0.5\textwidth]{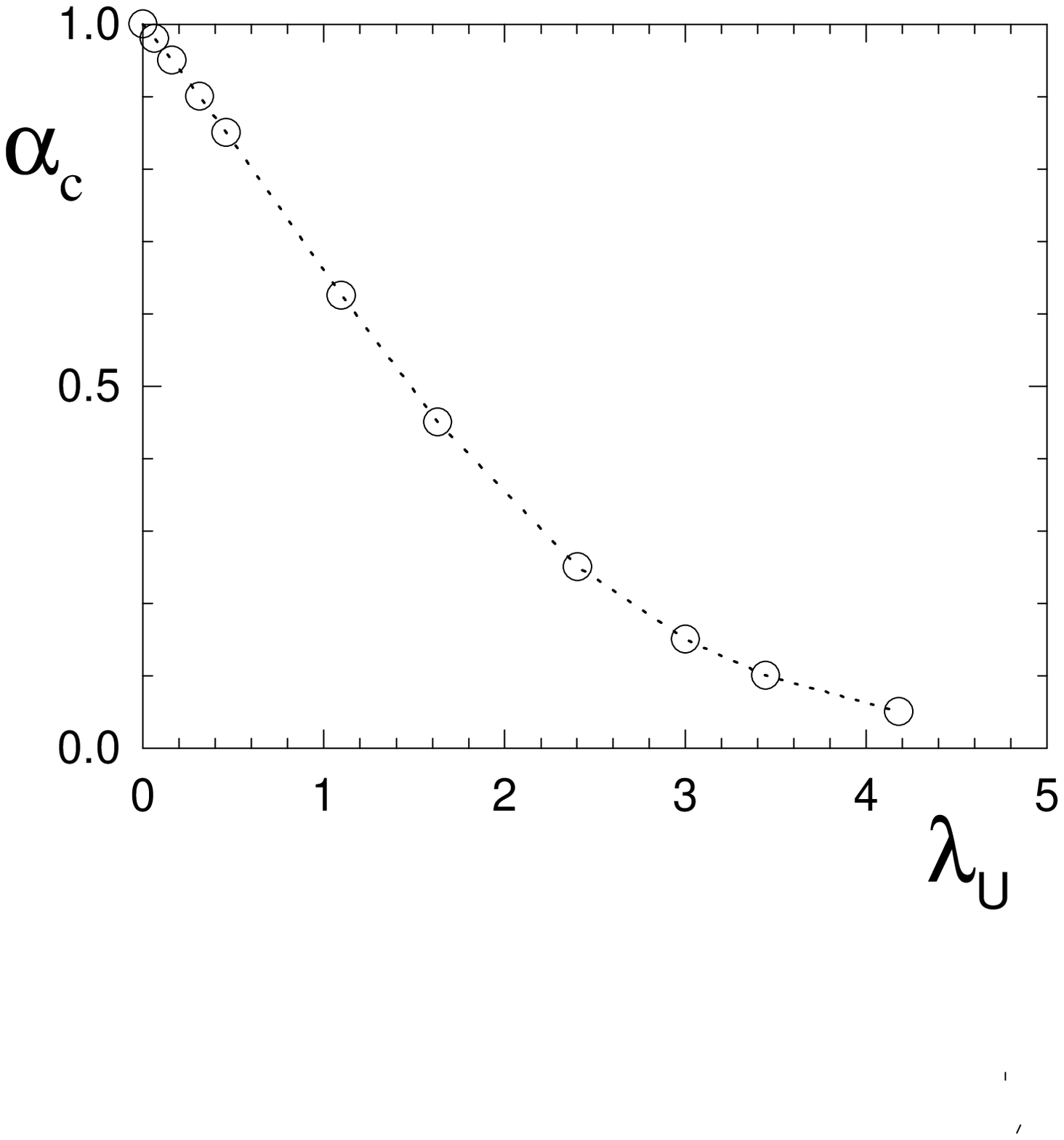}
\end{center}
\vspace*{-5mm}
\caption{\protect Critical value $\alpha_c$, below which synchronization 
is stable even in the thermodynamic limit, as a function of $\lambda_U$ 
(symbols), determined from Eq. (\ref{alfac}). 
The dotted line is a guide to the  eyes. 
} 
\label{fig:alfacrit} 
\end{figure}

If the positive Lyapunov exponent of the uncoupled map increases, the 
synchronization domain shrinks, collapsing in the limit $\lambda_U\to\infty$. 
In the opposite limit of $\lambda_U\to 0^+$ one gets

\begin{eqnarray} 
\label{crit1} 
\varepsilon_c &\to&  0 \hspace*{2cm}\mbox{and} \\ 
\label{crit2}
\varepsilon^\prime_c  &\to&  
 2 \biggl(1- \frac{b^{(N')}}{\eta}\biggr)^{-1} .
\end{eqnarray} 
This limit value of $\varepsilon^\prime_c$ depends on $\alpha$ and $N$. 
If $N\to\infty$ and $\alpha\to 0 (\infty)$, 
$\varepsilon^\prime_c$ goes to 2.0 (1.0) in the limit of vanishing chaos. 
(Although $N=21$, all these extreme behaviors are already 
insinuated in Fig.~\ref{fig:chaoticity}.) 
As a consequence, the critical value $\alpha_c<1$ below which 
the lattice synchronizes in the thermodynamic limit depends on the 
degree of chaoticity of the 
uncoupled maps. This dependence can be explicitly obtained by inversion of 

\begin{equation} \label{alfac}
\lambda_U =  \ln\left[ \frac{2}{C(\alpha_c)}-1\right]
\end{equation}
extracted from Eqs.~(\ref{e1inf}) and (\ref{e2inf}). 
The critical value $\alpha_c$ as a function of $\lambda_U$  
is displayed in Fig.~\ref{fig:alfacrit}. 
In the limit of vanishing (infinite) $\lambda_U$, $\alpha_c$ goes 1.0 (0.0).

%\vspace*{5mm}

\begin{figure}[htb] 
\begin{center} 
\includegraphics*[bb=40 180 560 590, width=0.5\textwidth]{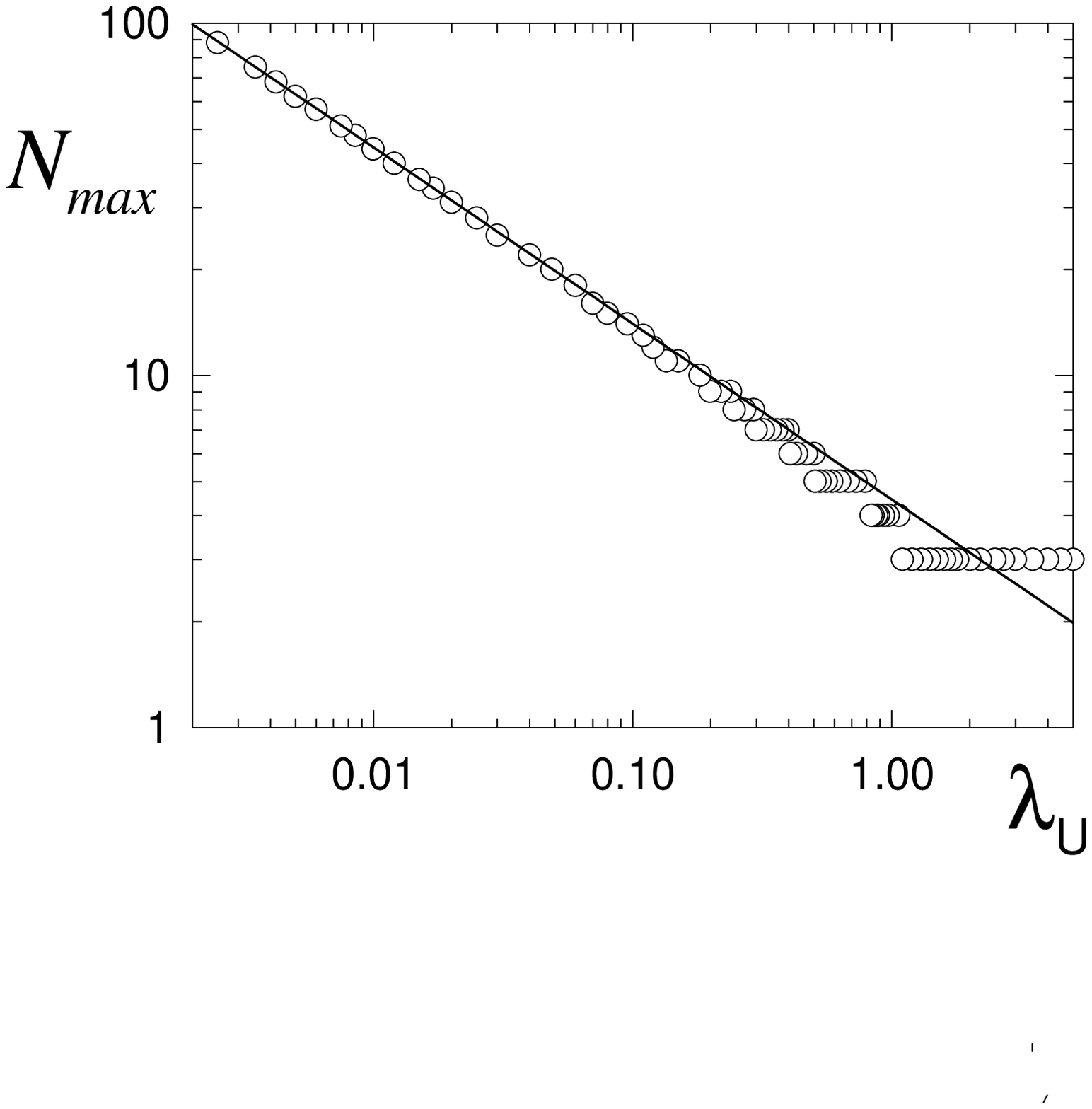}
\end{center} 
\vspace{-5mm} 
\caption{\protect Maximal lattice size $N_{max}\equiv N_b(\alpha,\lambda_U)$, 
for which synchronization 
can be achieved for any $\alpha$, as a function of the chaoticity indicator 
$\lambda_U$ (circles), determined from the condition 
$\varepsilon^\prime_c>\varepsilon_c$ for $\alpha=\infty$.
The solid line corresponds to the approximation given by Eq.~(\ref{Ncrit}).
} 
\label{fig:nmax} 
\end{figure}

When $\alpha>\alpha_c$, it must be $N\leq N_b(\alpha,\lambda_U)$ 
for the lattice to synchronize, where $N_b$  decreases with  
increasing $\alpha-\alpha_c$ (as shown in Fig~\ref{fig:Ne}). 
In the limit $\alpha\to\infty$, it is easy to obtain, 
from the condition $\varepsilon^\prime_c > \varepsilon_c$, 
an approximate expression for the maximal size, valid when 
one has sufficiently small $\lambda_U$ and large $N_b$:

\begin{equation} \label{Ncrit}
N_{max} \equiv N_b(\infty,\lambda_U) \;\leq\; \pi\sqrt{\frac{2}{\lambda_U}}   \;.
\end{equation}
In Fig.~\ref{fig:nmax}, we exhibit the maximal size $N_{max}$ for which synchronization 
can be achieved  in the limit of nearest-neighbor couplings (hence for any $\alpha$)  
as a function of $\lambda_U$, together with the approximation given by Eq. (\ref{Ncrit}). 

\vspace{5mm}

%\section*{Summary}

Summarizing, we have presented numerical and analytical results for the CS 
states in lattices of coupled identical chaotic maps with interactions that 
decay with distance as a power law. 
We have scrutinized the role of the system parameters 
in the ability of the lattice to attain complete synchronization using 
various chaotic maps. 
We observed, in the coupling parameters plane, an overall 
decrease of the area of the synchronization regions, as the number of 
coupled maps is increased. 
Moreover, the shape of these regions is bounded by critical curves which 
vary with the lattice size in a fashion we were able to predict analytically 
based on the Lyapunov spectrum of the synchronized state of the lattice, 
for virtually any chaotic map, with excellent agreement with numerical results. 
Similar findings have been given for the dependence of the synchronization regions 
on the degree of chaoticity of uncoupled maps, as well as for the maximal lattice 
size for which complete synchronization of chaos can be achieved.

\section*{Acknowledgments:}
We thank Sandro E. de S. Pinto for interesting discussions. 
This work was partially supported by Brazilian agencies CNPq, FAPERJ,  
Funda\c{c}\~ao Arauc\'aria and PRONEX.

\end{multicols}
\end{document}